\documentclass[12pt]{iopart}
\pdfoutput=1
\usepackage{epsfig}
\usepackage{amssymb,amsfonts}
\usepackage{mathrsfs}
\usepackage[bbgreekl]{mathbbol}
\usepackage{bbm}
\usepackage{slashed}
\usepackage{graphicx}
\usepackage{verbatim}
\usepackage[T1]{fontenc}
\usepackage[utf8]{inputenc}
\usepackage{ifthen}
\usepackage{forloop}
\usepackage[english]{babel}
\usepackage[usenames]{color}
\usepackage[dvipsnames]{xcolor}
\usepackage{bm}
\usepackage[normalem]{ulem}
\usepackage{footmisc}
\usepackage{cite}

\definecolor{darkgreen}{rgb}{0.2,0.6,0}
\definecolor{lightblue}{rgb}{0,0.5,0.8}
\definecolor{lightred}{rgb}{0.8,0.2,0.2}
\definecolor{darkorange}{rgb}{1,0.549,0}

\usepackage[colorlinks, pagebackref, citecolor=darkgreen]{hyperref}

\newcommand{\FRG}{{\small FRG}}

\newcommand{\eg}{{\textit{e.g.}}}
\newcommand{\ie}{{\textit{i.e.}}}

\newcommand{\be}{\begin{equation}}
\newcommand{\ee}{\end{equation}}
\newcommand{\bea}{\begin{eqnarray}}
\newcommand{\eea}{\end{eqnarray}}

\begin{document}

\title{Lessons from conformally reduced quantum gravity}

\author{Benjamin Knorr\,\href{https://orcid.org/0000-0001-6700-6501}{\protect \includegraphics[scale=.07]{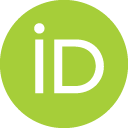}}}

\address{Perimeter Institute for Theoretical Physics, 31 Caroline St. N., Waterloo, ON N2L 2Y5, Canada}

\begin{abstract}
In this work we study a significantly enlarged truncation of conformally reduced quantum gravity in the context of Asymptotic Safety, including all operators that can be resolved in such a truncation including up to the sixth order in derivatives. A fixed point analysis suggests that there is no asymptotically safe fixed point in this system once one goes beyond an Einstein-Hilbert approximation. We will put these findings into context and discuss some lessons that can be learned from these results for general non-perturbative renormalisation group flows.
\end{abstract}
\maketitle

\section{Introduction}

One of the longest standing open puzzles of modern theoretical physics is how to quantise gravity. Despite decades of effort of many physicists around the world, it can be safely said that no universally accepted and experimentally verified theory of quantum gravity exists to date. Several proposals are on the market, each with their own advantages and disadvantages. One of those proposals is the so-called Asymptotic Safety Scenario \cite{Weinberg:1980gg, Reuter:1996cp}, which stipulates that gravity can be defined non-perturbatively by an interacting renormalisation group fixed point, for modern reviews see \cite{Percacci:2017fkn, Reuter:2019byg, Pawlowski:2020qer, Reichert:2020mja}. Clear advantages of this approach are its formulation in terms of standard quantum field theory concepts and its straightforward inclusion of matter \cite{Dona:2013qba, Dona:2015tnf, Meibohm:2015twa, Christiansen:2017cxa, Eichhorn:2018akn, Eichhorn:2018yfc}, and several milestones have been achieved in the past years. A key result is that the relevance of a given operator seems to be rather close to its canonical mass dimension even at the interacting fixed point \cite{Falls:2013bv, Falls:2014tra, Falls:2018ylp, Kluth:2020bdv}, which is good news for setting up reliable approximation schemes. Moreover, the inclusion of the two-loop counterterm keeps the fixed point intact \cite{Gies:2016con}. Recently, a lot of progress has been made in the resolution and investigation of the convergence of momentum-dependent correlation functions \cite{Christiansen:2014raa, Christiansen:2015rva, Meibohm:2015twa, Denz:2016qks, Christiansen:2017cxa, Eichhorn:2018ydy, Pawlowski:2020qer}, and their inclusion in the calculation of observables like scattering cross sections \cite{Draper:2020bop, Draper:2020knh}. Some of the programs open problems \cite{Donoghue:2019clr, Bonanno:2020bil} are the question about a Lorentzian formulation \cite{Manrique:2011jc, Rechenberger:2012dt, Biemans:2016rvp, Biemans:2017zca, Houthoff:2017oam, Knorr:2018fdu}, the proper implementation of background independence \cite{Pawlowski:2005xe, Becker:2014qya, Christiansen:2014raa, Christiansen:2015rva, Dietz:2015owa, Denz:2016qks, Labus:2016lkh, Morris:2016spn, Morris:2016nda, Percacci:2016arh, Safari:2016gtj, Safari:2016dwj, Christiansen:2017bsy, Lippoldt:2018wvi} and the conceptual understanding of the convergence properties of approximation schemes \cite{Denz:2016qks, Eichhorn:2018akn}.

In all these calculations, it is expected that the dynamics of the renormalisation group flow is dominantly driven by the physical (spin two) transverse traceless fluctuations. Notwithstanding this expectation, the fixed point used for the definition of the theory seems also to be present in a dramatically simplified model, the so-called conformally reduced approximation. In this truncation, only fluctuations of the conformal mode are integrated out, all other degrees of freedom (including the physical spin two part) are frozen. The big advantage of this setup is the much lower technical complexity, since one doesn't need a gauge fixing, and effectively, one studies a special scalar theory in curved spacetime. In an Einstein-Hilbert truncation, the fixed point of this setup shows characteristics similar to that of the ``full theory'', where all fluctuations are integrated out \cite{Reuter:2008wj, Reuter:2008qx}. Upon inclusion of an $R^2$-term, the fixed point only exists if matter is included \cite{Machado:2009ph}. The fixed point persists upon the inclusion of a form factor interaction involving the Weyl tensor \cite{Bosma:2019aiu}. The full field dependence on the conformal factor was discussed in \cite{Dietz:2015owa}. Results on conformally reduced $f(R)$ gravity can be found in \cite{Demmel:2012ub, Demmel:2013myx, Demmel:2014sga} for three dimensions, and \cite{Demmel:2015oqa} for four dimensions.

An independent argument for why the conformal mode might be important in the quantum regime of gravity is the conformal factor instability \cite{Gibbons:1978ac, Mazur:1989by, Mottola:1995sj, Dasgupta:2001ue}. It stems from the fact that the Einstein-Hilbert action is unbounded from below. This is related to the kinetic operator of the conformal factor having the ``wrong'' sign. It has been argued that higher order derivative terms might cure this problem \cite{Antoniadis:1991fa, Machado:2009ph}. For early work on conformally reduced gravity, see also \cite{Antoniadis:1992xu, Antoniadis:1992pr, Antoniadis:1995dx, Antoniadis:1995dy, Antoniadis:1996pb, Mazur:2001aa}.

These results give rise to the following question: is the dynamics of quantum gravity driven by the conformal mode? Or are the fixed point results an artefact of the approximations of restricting the action to a certain subset of operators? We will investigate these questions in this work by studying the non-perturbative renormalisation group flow of the conformally reduced model, including all operators with up to six derivatives acting on the metric. The results indicate that the conformally reduced fixed point in the Einstein-Hilbert truncation is indeed a truncation artefact. Assuming that quantum gravity is asymptotically safe, the renormalisation group flow of full quantum gravity is then indeed dominantly driven by the spin two sector.

This work is structured as follows: in section \ref{sec:FRG} we quickly review the tool that we use to investigate non-perturbative renormalisation group flows, the functional renormalisation group, and define the notion of Asymptotic Safety. Following that, in section \ref{sec:ansatz} we specify the details and assumptions of our approximation. Section \ref{sec:RGflow} contains a discussion on the derivation of the renormalisation group flow, followed by a step-by-step fixed point analysis in section \ref{sec:FPanalysis}. Afterwards, in section \ref{sec:derivexp} we discuss the shortcomings of our approximation, and more generally any approximation based on the derivative expansion.
We close with a discussion of the results in section \ref{sec:discussion}. \ref{app:beta} contains the explicit form of all renormalisation group equations, and \ref{app:basis} provides a basis for operators with eight derivatives for future reference.

\section{Functional Renormalisation Group and Asymptotic Safety}\label{sec:FRG}

The most commonly used tool to investigate the non-perturbative renormalisation group flow of quantum gravity in the continuum is the functional renormalisation group (\FRG{}).\footnote{A closely related approach to study Asymptotic Safety is the Causal Dynamical Triangulation program \cite{Ambjorn:2012jv, Loll:2019rdj}, which is a lattice formulation of the quantum gravity path integral.} Its central object of study is the effective average action $\Gamma_k$, which interpolates between the classical action $S$ at $k=\infty$ and the standard quantum effective action $\Gamma$ at $k=0$. Its dependence on the scale $k$ is governed by the Wetterich equation \cite{Wetterich:1992yh, Morris:1993qb},
\begin{equation}\label{eq:flow}
 k \partial_k \Gamma_k = \frac{1}{2} {\rm{STr}} \left[ \frac{1}{\Gamma_k^{(2)} + \mathfrak R_k} k \partial_k \mathfrak R_k \right] \, .
\end{equation}
In this, $\mathfrak R_k$ is a scale-dependent regulator, $\Gamma^{(2)}_k$ is the second variation of the effective average action with respect to all fluctuating fields, and STr combines a functional trace over continuous indices with a sum over discrete indices. The equation can be derived by considering a slightly modified Legendre transformation of the partition function. For reviews on the \FRG{} see \cite{Berges:2002ew, Pawlowski:2005xe, Gies:2006wv, Delamotte:2007pf, Dupuis:2020fhh}.

Given an ansatz for $\Gamma_k$ and a complete operator basis, both sides of the flow equation \eref{eq:flow} can be evaluated and the beta functions $\beta_{g_i}$ of the $k$-dependent couplings $g_i$ can be read off. The beta functions are the differential equations that govern the scale dependence of the couplings of the theory in units of the scale $k$, \ie{} one considers the dimensionless counterparts of the couplings. For example, for the cosmological constant $\Lambda$ we consider the dimensionless coupling $\lambda$ and its beta function $\beta_\lambda$ defined by
\begin{equation}
 \lambda = \Lambda k^{-2} \, , \qquad \beta_\lambda \equiv k \partial_k \lambda \equiv \dot \lambda \, .
\end{equation}
Here we introduced the shorthand $\dot x = k \partial_k x$. A fixed point is a combined zero of all beta functions. If not all couplings at a fixed point are zero, we speak of an interacting, or asymptotically safe fixed point. To learn about the stability properties of a fixed point, one can linearise the flow around it and study the eigenvalues of the stability matrix. Negative eigenvalues indicate that the renormalisation group flow in the direction of the eigenvector is attracted towards the fixed point when considering the flow pointing towards larger $k$, that is the ultraviolet. The corresponding couplings are called relevant. Likewise, positive eigenvalues correspond to repulsive directions, and are called irrelevant. Predictive fixed points are those which only have a finite number of relevant couplings. The number of the relevant couplings corresponds to the number of independent experiments that have to be conducted to uniquely fix the theory.

The \FRG{} has been applied to quantum gravity, and in practically all studies so far, a suitable predictive fixed point (the so-called Reuter fixed point \cite{Reuter:1996cp}) has been found that allows the non-perturbative renormalisation of the theory.

\section{Ansatz for the effective action and regularisation}\label{sec:ansatz}

In practice, we usually cannot solve the flow equation \eref{eq:flow} exactly, but have to resort to approximations. We will now specify the approximation that we employ in this work.
Our ansatz for the effective average action includes all operators with up to 6 derivatives acting on the metric which can be disentangled in a conformally reduced setup. This entails
\begin{equation}\label{eq:actionansatz}
 \Gamma_k = \frac{1}{16\pi G_N} \int \rmd^4x \sqrt{g} \left( \mathcal L_0 + \mathcal L_2 + \mathcal L_4 + \mathcal L_6 \right) \, ,
\end{equation}
where the $\mathcal L_{2i}$ are the parts of the action with $2i$ derivatives, and $G_N$ is Newton's constant. Our choice of basis for the individual pieces reads
\begin{equation}
 \mathcal L_0 = 2\Lambda \, , \qquad
 \mathcal L_2 = -R \, , \qquad
 \mathcal L_4 = -\frac{1}{6} g_{R^2} R^2 + \frac{1}{2} g_{C^2} C^{\mu\nu\rho\sigma} C_{\mu\nu\rho\sigma} \, ,
\end{equation}
and
\begin{eqnarray}
 \mathcal L_6 &= -\frac{1}{6} g_{R\Delta R} R \Delta R + \frac{1}{2} g_{C\Delta C} C^{\mu\nu\rho\sigma} \Delta C_{\mu\nu\rho\sigma}+ g_{R^3} R^3  \nonumber \\
 &\quad+ g_{RS^2} R S^{\mu\nu} S_{\mu\nu} + g_{S^3} S^\mu_{\phantom{\mu}\nu} S^\nu_{\phantom{\nu}\rho} S^\rho_{\phantom{\rho}\mu}+ g_{SSC} S^{\mu\nu} S^{\rho\sigma} C_{\mu\rho\nu\sigma} \nonumber \\
 &\quad  + g_{RC^2} R C^{\mu\nu\rho\sigma} C_{\mu\nu\rho\sigma}+ g_{C^3} C^{\mu\nu}_{\phantom{\mu\nu}\rho\sigma} C^{\rho\sigma}_{\phantom{\rho\sigma}\tau\omega} C^{\tau\omega}_{\phantom{\tau\omega}\mu\nu} \, .
\end{eqnarray}
Signs and prefactors are chosen to simplify the form of the propagator. For future reference, we also present a full basis for the operators of the next order in the derivative expansion in \ref{app:basis}. All couplings are understood to depend on the renormalisation group scale $k$.

In our choice of operators we are using a trace-free basis with the trace-free Ricci tensor $S$ defined by
\begin{equation}
 S_{\mu\nu} = R_{\mu\nu} - \frac{1}{4} R g_{\mu\nu} \, ,
\end{equation}
and the Weyl tensor $C$ given by
\begin{equation}
 C^{\mu\nu}_{\phantom{\mu\nu}\rho\sigma} = R^{\mu\nu}_{\phantom{\mu\nu}\rho\sigma} - 2 R^{[\mu}_{[\rho} \delta^{\nu]}_{\sigma]} + \frac{1}{3} R \delta^{[\mu}_{\rho} \delta^{\nu]}_\sigma \, .
\end{equation}
The only operator which cannot be resolved in a conformally reduced setup up to this order is the $C^2$-operator, since it is conformally invariant. For that reason, its coupling constant $g_{C^2}$ will not appear on the right-hand side of the flow equation. Correspondingly we will not ask for a fixed point in this coupling since otherwise the system is generically over-constrained (see however the discussion in section \ref{sec:derivexp}), and set it to zero throughout.

To define a clean projection for the flow of $g_{R^2}$, we complete the basis by including the (density of the) Euler characteristic $\mathfrak E$,
\begin{equation}
 \mathfrak E = -C^{\mu\nu\rho\sigma} C_{\mu\nu\rho\sigma} + 2S^{\mu\nu} S_{\mu\nu} - \frac{1}{6} R^2 \, ,
\end{equation}
which is a topological invariant in four dimensions. In this way the complete set of operators with four derivatives reads
\begin{equation}
 \int \rmd^4x \sqrt{g} \left\{ R^2, C^2, \mathfrak E \right\} \, .
\end{equation}
Our prescription is then to express the right-hand side of the flow equation in this basis, and map the prefactor of the $R^2$-term to the flow of $g_{R^2}$. Explicitly, this map reads
\begin{eqnarray}
 \int \rmd^4x \sqrt{g} \left[ z_R R^2 + z_S S^2 + z_C C^2 \right] \nonumber \\
 \mapsto \int \rmd^4x \sqrt{g} \left[ \left( z_R + \frac{z_S}{12} \right) R^2 + \left( z_C + \frac{z_S}{2} \right) C^2 - \frac{z_S}{2} \mathfrak E  \right] \, .
\end{eqnarray}
This entails that the flow of $g_{R^2}$ involves the traces proportional to $R^2$ and $S^2$.

To calculate the second variation of $\Gamma_k$, we have to specify how we parameterise the conformal fluctuations of the metric. We do so by an exponential parameterisation,
\begin{equation}
 g_{\mu\nu} = e^{\phi} \bar g_{\mu\nu} \, ,
\end{equation}
Here $g$ is the full metric, $\bar g$ is an arbitrary background metric and $\phi$ is the conformal fluctuation. All variations are performed with respect to $\phi$, and once we have evaluated the second variation, we set the fluctuation to zero (background field approximation). For results in conformally reduced gravity beyond such an approximation see \cite{Dietz:2015owa}.

It is left to define our regularisation scheme. We opt for a regulator without any endomorphism term (sometimes referred to as ``type I'' regulator \cite{Codello:2008vh}), so that the two-point function of the conformal factor reads structurally
\begin{equation}\label{eq:2ptfct}
 \Gamma_k^{(2)}+\mathfrak R_k \propto \frac{1}{G_N} \left( \Delta + g_{R^2} \Delta^2 + g_{R\Delta R} \Delta^3 + R_k(\Delta) - \frac{8}{3}\Lambda \right) + \dots \, ,
\end{equation}
where the dots stand for the suppressed curvature terms.
For the explicit form of the shape function $R_k$, we chose an exponential,
\begin{equation}
 R_k(z) = \frac{z}{e^{\frac{z}{k^2}}-1} \, .
\end{equation}
We checked a few other shape functions to ensure that all results are qualitatively stable under regulator changes.

\section{Conformally reduced renormalisation group flow}\label{sec:RGflow}

Now we are in the situation to derive the flow equations. Since our action is based on a derivative expansion, we will expand everything on the right-hand side in curvatures and derivatives thereof. Practically, this works in the following way. We first note that from \eref{eq:2ptfct} we have the structural form
\begin{equation}
 \Gamma^{(2)}+\mathcal R_k = \mathcal P(\Delta) + \mathcal F \, ,
\end{equation}
where $\mathcal P$ is a function of the Laplacian only, and the operator $\mathcal F$ contains all terms that have at least one factor of a curvature. To calculate the renormalisation group flow, we have to invert this object. Since we are interested in an expansion in curvatures, we can perform the inversion with a truncated geometric series,
\begin{equation}
 \mathfrak G = \left( \Gamma^{(2)}+R_k \right)^{-1} = \sum_{\ell\geq0} \left( - \mathcal P^{-1} \mathcal F \right)^\ell \mathcal P^{-1} \equiv \sum_{\ell\geq0} \left( - G \mathcal F \right)^\ell G \, .
\end{equation}
The two points that make this calculation straightforward is that since $\mathcal P$ is a function of the Laplacian $\Delta$ only, its inverse $G=\mathcal P^{-1}$ is easily calculated without any need to involve commutators. Second, since we are interested in a finite order of the expansion, only terms with $\ell\leq3$ contribute to the functional trace. This entails that
\begin{eqnarray}
 \dot \Gamma_k &\simeq \frac{1}{2} {\rm{Tr}} \bigg[ \left( G - G \mathcal F G + G \mathcal F G \mathcal F G - G \mathcal F G \mathcal F G \mathcal F G \right) \dot \mathfrak R_k \bigg] \nonumber \\
 &= \frac{1}{2} {\rm{Tr}} \bigg[ \left( 1 - \mathcal F G + \mathcal F G \mathcal F G - \mathcal F G \mathcal F G \mathcal F G \right) G \, \dot \mathfrak R_k \bigg] \nonumber \\
 &\simeq \frac{1}{2} {\rm{Tr}} \bigg[ \left( 1 - \mathcal F G + \mathcal F [G, \mathcal F] G + \mathcal F^2 G^2 - \mathcal F^3 G^3 \right) G \, \dot \mathfrak R_k \bigg] \, .
\end{eqnarray}
The symbol $\simeq$ indicates equality up to terms that don't contribute in the considered truncation. In the first step, we used the cyclicity of the trace to sort the left-most factor of $G$ to the right. Since by our choice \eref{eq:2ptfct}, $\mathfrak R_k$ is also a pure function of $\Delta$, the two commute. In the second step, we sorted all remaining factors of $G$ to the right, and only kept terms which contribute to cubic order. The sorting of all functions of the Laplacian to the very right is necessary to straightforwardly employ heat kernel techniques. The single commutator that is left can be calculated by using an intermediate inverse Laplace transform and the Baker-Campbell-Hausdorff formula \cite{Benedetti:2010nr, Knorr:2019atm}:
\begin{eqnarray}
 [f(\Delta),\mathcal O] &= \int_0^\infty {\rm{d}}s \, \tilde f(s) \, \left[ e^{-s\Delta}, \mathcal O \right] \nonumber \\
 &= \int_0^\infty {\rm{d}}s \, \tilde f(s) \sum_{n\geq1} \frac{(-s)^n}{n!} \left[ \Delta, \mathcal O \right]_n e^{-s\Delta} \nonumber \\
 &= \sum_{n\geq 1} \frac{1}{n!} \left[ \Delta, \mathcal O \right]_n \, f^{(n)}(\Delta) \, .
\end{eqnarray}
Here, we used the nested commutator
\begin{equation}
 [X,Y]_1 = [X,Y] \, , \qquad [X,Y]_{n+1} = [X,[X,Y]_n] \, .
\end{equation}
With this, we can write the expression for the renormalisation group flow as
\begin{equation}
\fl  \dot \Gamma_k \simeq \frac{1}{2} {\rm{Tr}} \bigg[ \left( 1 - \mathcal F G + \mathcal F [\Delta, \mathcal F] G' G + \frac{1}{2} \mathcal F [\Delta, \mathcal F]_2 G'' G + \mathcal F^2 G^2 - \mathcal F^3 G^3 \right) G \, \dot \mathfrak R_k \bigg] \, .
\end{equation}
The primes indicate derivatives w.r.t.~$\Delta$. Once again we neglected terms that don't contribute to our approximation. For this it is crucial to realise that each commutator increases the mass dimension of the operator insertion by at least one. The striking simplicity of the expression explains why the conformally reduced setup is technically so much simpler: the only non-trivial tensor structure appears inside of $\mathcal F$ with all indices completely contracted.

The expression in brackets is now in a form which can be straightforwardly evaluated with heat kernel techniques \cite{Vassilevich:2003xt, Groh:2011dw}.\footnote{The necessary heat kernel coefficients have been recalculated, and a minor typographical error in the heat kernel coefficient $\overline{A_3}$ quoted in \cite{Groh:2011dw} has been accounted for. The relevant change is the prefactor of the term $R \Delta R$, whose correct value is $-\frac{1}{180}$ instead of $-\frac{1}{280}$, in agreement with \cite{Barvinsky:1993en}.} The result is given in \ref{app:beta}, and has been obtained with the help of the \emph{xAct} suite for Mathematica \cite{xActwebpage, 2007CoPhC.177..640M, Brizuela:2008ra, 2008CoPhC.179..597M, 2014CoPhC.185.1719N}.

\section{Fixed point analysis}\label{sec:FPanalysis}

Having derived the renormalisation group equations for the ansatz \eref{eq:actionansatz}, we will now perform a fixed point analysis. We will do this step by step, starting with an Einstein-Hilbert truncation, then including the $R^2$-term, followed by including $R^3$ and finally discussing the full setup.

\paragraph{General properties}

Before we dive into the different levels of approximation, we will point out some general properties of the renormalisation group flow. The first finding is that the flow of the identity operator \eref{eq:flow1}, corresponding to the renormalisation group flow of the cosmological constant, is strictly positive. This has the profound consequence that, at the fixed point, we need
\begin{equation}
 \frac{\lambda}{g} > 0 \, ,
\end{equation}
where $g$ and $\lambda$ are the dimensionless Newton's and cosmological constant, respectively. This implies that, since we need a positive Newton's constant, that also the cosmological constant at the fixed point must be positive. Combining this with the condition
\begin{equation}
 \lambda < \frac{3}{8} \, ,
\end{equation}
which comes from the pole condition of the propagator, we only have to investigate the strip
\begin{equation}
 0 < \lambda < \frac{3}{8} \, ,
\end{equation}
to search for physically interesting fixed points. This result is independent of the level of truncation in the conformally reduced model, since it only relies on positivity properties of the regulator and the propagator.

The second observation that we want to make is the functional form of the renormalisation group equations. By the nature of the derivative expansion, the right-hand side of the flow equation will depend rationally on all couplings that contribute to the flat propagator. In particular, this includes the cosmological constant and all couplings related to operators of the form $R \Delta^n R$. Terms of the same form with Weyl tensors instead of the Ricci scalars will not contribute to the conformal propagator, but would contribute to the spin-two graviton propagator in calculations which take all fluctuations into account. All other couplings will appear polynomially in each individual beta function. This allows to solve a subset of the beta functions for a partial fixed point analytically which reduces the numerical workload significantly. As an exception to this rule, since we introduced Newton's constant as a global prefactor of the whole action, we can solve its beta function for its unique non-vanishing fixed point value, even though it technically appears in the flat propagator.

\paragraph{Einstein-Hilbert truncation}

The Einstein-Hilbert truncation is the most common approximation scheme in the literature on Asymptotic Safety. It retains the cosmological constant and Newton's constant, which is equivalent to truncating the derivative expansion at second order. In this truncation, we find a single fixed point at
\begin{equation}
 g = 1.23 \, , \qquad \lambda = 0.371 \, .
\end{equation}
The fixed point is situated quite close to the singular line $\lambda = 3/8$, which is in agreement with previous studies \cite{Reuter:2008wj, Reuter:2008qx, Machado:2009ph} when factoring in small differences due to different regularisation choices.

\paragraph{$R^2$-truncation}

The next approximation level that we will study is the fourth order derivative expansion, which retains the $R^2$ operator in addition to the operators of the Einstein-Hilbert truncation. In this truncation, we have to solve three fixed point equations. Once again, we can solve the equation for the running Newton's constant analytically, and then study the zeros of the two remaining beta functions. It turns out that the zeros of these beta functions don't intersect, which means that there is no fixed point in the physical regime. This is in agreement with a previous study of this truncation in the conformally reduced setting \cite{Machado:2009ph}, where it was also found that the inclusion of matter might reintroduce the fixed point. A point to consider is that in truncations of $f(R)$-type in full quantum gravity, it has been found that this level of truncation seems to be anomalously imprecise, and the inclusion of higher order operators could stabilise the system again \cite{Codello:2007bd, Machado:2007ea, Falls:2013bv, Falls:2014tra, Falls:2018ylp, Kluth:2020bdv}. With this in mind, we will now include sixth order derivative terms.

\paragraph{$R^3$-truncation}

As an intermediate step and to make contact with $f(R)$-type truncations \cite{Demmel:2012ub, Demmel:2013myx, Demmel:2014sga, Demmel:2015oqa}, we will first only add the operator $R^3$. Within this truncation, we can solve for Newton's coupling again. Plugging in its value into the beta function of the cosmological constant, we can in fact solve this equation uniquely for $g_{R^3}$, and we are left with finding zeros of the beta functions for the $R^2$ and $R^3$ couplings in terms of the cosmological constant and $g_{R^2}$. A numerical analysis shows however again that no fixed point exists in the physical regime. On the solution of the previous equations, the latter two beta functions are strictly positive in the allowed range for $\lambda$. This is to be contrasted with the results in \cite{Demmel:2015oqa}, where a fixed point in an $f(R)$ truncation was found, suggesting that this order of the approximation might be imprecise as well.

\paragraph{Full sixth order truncation}

Finally, we will discuss the full truncation with all sixth order operators. Once again, we can solve several beta functions analytically to eliminate some of the couplings. In particular, we can obtain the fixed point values of the couplings $g, g_{C\Delta C}, g_{C^3}, g_{RC^2}, g_{SSC}$ and $g_{R^3}$, whereas we have a cubic equation for $g_{S^3}$.\footnote{In all cases that were probed numerically, only one of the three roots is real.} With this, it remains to find a combined zero of the beta functions for the couplings $g_{R^2}, g_{R^3}, g_{RS^2}$ and $g_{R\Delta R}$ in terms of the couplings that enter the flat spacetime propagator, $\lambda, g_{R^2}, g_{R\Delta R}$, and the coupling $g_{RS^2}$. We have performed an extensive numerical fixed point search, but haven't found any fixed point in the physical region.

\section{Shortcomings of the derivative expansion}\label{sec:derivexp}

In this section we will discuss two issues of the derivative expansion related to the (non-)existence of fixed points in a finite order derivative expansion, which are not specific to gravity or our model.

First, in such an expansion, the regularised propagator has the form
\begin{equation}\label{eq:DEprop}
 G_{\rm{DE}}(p^2) \propto \frac{1}{p^2 + g_4 p^4 + g_6 p^6 + \dots + R_k(p^2) + \mu} \, ,
\end{equation}
where for the sake of the argument we assumed a flat (Euclidean) background so that we can perform a Fourier transform, and neglected any potential tensor structures. Here, $\mu$ is a mass term and the $g_i$ are coupling constants corresponding to $2i$-derivative monomials. Propagators with the loop momentum as their argument will be integrated over in the flow equation. From this it follows that in a finite order truncation, the existence of a fixed point depends strongly on the sign of the highest $g_i$ that is retained in a given approximation. In particular, we will only find a fixed point if the highest coefficient is positive\footnote{Let us mention here that a potential overall minus sign is already assumed to be accounted for in \eref{eq:DEprop}, which is, \eg{}, necessary to define a reasonable regularisation of the conformal mode.},
\begin{equation}
 g_{i_{\rm{max}}} > 0 \, .
\end{equation}
Otherwise the propagator has a pole at finite momentum, and the renormalisation group flow is ill-defined. On the other hand, assume that we have the fully momentum-dependent propagator, and perform a Taylor expansion of it. Generically, we expect that the series coefficients can have either sign. If our truncation happens to be at an order where the highest coefficient is negative, we will not see the fixed point at this order, because the renormalisation group flow is ill-defined in this region of theory space.\footnote{This assumes that the coefficient has the same sign in the truncation as in the exact solution.} Clearly this is then an artefact of the finite order of the expansion. From this we see that from the non-existence of a fixed point at a certain order of the derivative expansion we cannot conclude that no fixed point exists - either several orders have to be investigated, or a fully momentum dependent calculation has to be performed. On the other hand, if a fixed point is found in a derivative expansion, this is not a proof that the fixed point exists. Nevertheless something can be learned from the pattern of including higher and higher orders in the expansion, if a fully momentum dependent calculation is unfeasible. In particular, if all poles are away from the non-negative real axis, then the renormalisation group flow is well-defined and finite, and one could expect that any predictions made in such a setup are valid inside the radius of convergence of the expansion. In this context one should also be mindful if one employs a regulator with compact support like the popular linear regulator \cite{Litim:2001up}, since it can mask poles outside of its domain of support.

Second, there is also an ``inverse'' statement: the full system might have a fixed point even though no finite order of the derivative expansion can see it. This is the case for conformally reduced gravity where we amend the Einstein-Hilbert action with couplings of the form
\begin{equation}
 g_{C\Delta^nC} C^{\mu\nu\rho\sigma} \Delta^n C_{\mu\nu\rho\sigma} \, .
\end{equation}
At any finite order, since the coupling with $n=0$ is exactly marginal in the conformally reduced model, we will not find a fixed point for that coupling. Nevertheless, once the full form factor is included,
\begin{equation}
 C^{\mu\nu\rho\sigma} f_{CC}(\Delta) C_{\mu\nu\rho\sigma} \, ,
\end{equation}
a fixed point exists \cite{Bosma:2019aiu}. The reason for that is that in a derivative expansion, we have to find a combined zero of $n$ functions which only depend on $n-1$ couplings. In the limit of $n\to\infty$, this has a chance of not being over-constrained.

\section{Discussion}\label{sec:discussion}

We investigated the non-perturbative renormalisation group flow of conformally reduced quantum gravity in the derivative expansion. While at second order in derivatives, there is a fixed point suitable for asymptotic safety which shares some features with the Reuter fixed point, it disappears once higher order derivative terms are included. This is to be contrasted with results in an $f(R)$-truncation, where a fixed point exists \cite{Demmel:2015oqa}. There are at least two possibilities how to interpret this. Either the additional tensor structures considered in this work indeed prevent the conformally reduced system from having a fixed point, or higher order operators that are resolved with a full function $f(R)$ stabilise the system again.

We tentatively conclude that the expectation that quantum gravity is mainly driven, or at least substantially stabilised, by the physical spin two sector, is indeed justified. As a consequence, it might be useful to study ``spin two only'' truncations where only the gauge-invariant spin two sector is allowed to fluctuate in order to investigate more intricate approximation schemes.

We also discussed generic shortcomings of the derivative expansion. Because the two-point function in such an approximation is polynomial, the propagator necessarily develops additional poles. If these poles are on the positive part of the real line, the renormalisation group flow is ill-defined. The resummation of these terms into form factors can cure this problem, although at considerable technical cost. The inclusion of form factors might nevertheless be strictly necessary to find all fixed points of a given system. Concerning conformally reduced gravity, it is thus still conceivable that a fixed point exists in an even more elaborate truncation involving more form factors, but even in this simple model the technical complexity will be formidable.

\ack
I would like to thank Frank Saueressig for critical comments on the manuscript. This work is supported by Perimeter Institute for Theoretical Physics. Research at Perimeter Institute is supported in part by the Government of Canada through the Department of Innovation, Science and Industry Canada and by the Province of Ontario through the Ministry of Colleges and Universities.

\newpage

\appendix

\section{Explicit beta functions}\label{app:beta}

In this appendix we collect the result of the functional trace. We first define the propagator function
\begin{equation}
 G(z) = \frac{1}{z + g_{R^2} z^2 + g_{R\Delta R} z^3 + R_k(z) - \frac{8}{3}\lambda} \, ,
\end{equation}
and for convenience introduce the measure
\begin{equation}
 \mu(z) = \frac{1}{32\pi^2} \left( \left(4 - \frac{\dot g}{g} \right)R_k(z) - 2z R_k'(z) \right) {\rm{d}}z \, .
\end{equation}
Now we successively go through all tensor structures of interest. We indicate the corresponding tensor structure after a vertical dash. The result of the functional trace is
\begin{eqnarray}
 \fl\left. \dot \Gamma_k \right|_{\mathbbm 1} &= \int_0^\infty \mu(z) G(z) z \, , \label{eq:flow1} \\
 \fl\left. \dot \Gamma_k \right|_{R} &= \frac{1}{6} \int_0^\infty \mu(z) G(z) \nonumber \\
 \fl &\quad + \frac{1}{6} \int_0^\infty \mu(z) G(z)^2 z \bigg[-2+2g_{R^2}z+(108g_{R^3} + 4 g_{R\Delta R} + 3g_{RS^2})z^2 \bigg] \, , \\
 \fl\left. \dot \Gamma_k \right|_{R^2} &= \frac{1}{72} \mu(0) G(0) + \frac{1}{18} \int_0^\infty \mu(z) G(z)^2 \bigg[ -1+g_{R^2} z - (108g_{R^3} + 7 g_{RS^2} + g_{S^3}) z^2 \bigg] \nonumber \\
 \fl &\quad + \frac{1}{144} \int_0^\infty \mu(z) G(z)^3 \bigg[ 16z - 32g_{R^2}z^2 + 16 (g_{R^2}^2-4(27g_{R^3} + g_{R\Delta R})-3g_{RS^2}) z^2 \nonumber \\
 \fl & \quad + 16g_{R^2}(108g_{R^3}+4g_{R\Delta R}+3g_{RS^2}) z^4 + \bigg\{ 46656g_{R^3}^2 + 64g_{R\Delta R}^2 + 96g_{R\Delta R} g_{RS^2} \nonumber \\
 \fl & \quad + 52 g_{RS^2}^2 + 864g_{R^3}(4g_{R\Delta R}+3g_{RS^2}) + 8g_{RS^2} g_{S^3} + g_{S^3}^2 \bigg\} z^5 \bigg] \, , \\
 \fl\left. \dot \Gamma_k \right|_{R\Delta R} &= \frac{1}{90} \mu(0) G(0)^2 + \frac{17}{5040} \left( \mu'(0) G(0) + \mu(0) G'(0) \right) \nonumber \\
 \fl & \quad + \frac{1}{72} \int_0^\infty \mu(z) G(z)^2 \bigg[ -2g_{R^2} + (4g_{R\Delta R}-4g_{RS^2}-g_{S^3})z \bigg] \nonumber \\
 \fl & \quad + \frac{1}{576} \int_0^\infty \mu(z) G(z)^3 \bigg[ 16(g_{R^2}^2-12(36g_{R^3}+3g_{R\Delta R}+g_{RS^2}))z^2 \nonumber \\
 \fl & \quad + 32g_{R^2}(324g_{R^3}+22g_{R\Delta R}+13g_{RS^2}+g_{S^3})z^3 \nonumber \\
 \fl & \quad + \bigg\{ 559872g_{R^3}^2+32(44g_{R\Delta R}^2+69g_{R\Delta R}g_{RS^2}+38g_{RS^2}^2)  \nonumber \\
 \fl & \quad + 8g_{S^3}(18g_{R\Delta R} + 37g_{RS^2}) + 25g_{S^3}^2 + 3456g_{R^3}(17g_{R\Delta R}+13g_{RS^2}+g_{S^3}) \bigg\} z^4 \bigg] \nonumber \\
 \fl & \quad + \frac{1}{288} \int_0^\infty \mu(z) G(z)^4 \bigg[ -3g_{R\Delta R}\bigg\{ 559872g_{R^3}^2 + 768g_{R\Delta R}^2 \nonumber \\
 \fl & \quad + 1984 g_{R\Delta R}g_{RS^2} + 1456g_{RS^2}^2 + 208g_{R\Delta R}g_{S^3} + 356g_{RS^2}g_{S^3} + 25g_{S^3}^2 \nonumber \\
 \fl & \quad + 432g_{R^3}(96g_{R\Delta R}+124g_{RS^2}+13g_{S^3}) \bigg\} z^7 - 32z(1+R_k'(z)) \nonumber \\
 \fl & \quad + 16(g_{R^2}(-1+7R_k'(z))-2R_k''(z))z^2 \nonumber \\
 \fl & \quad + 16\bigg\{ 17g_{R^2}^2 + 540g_{R^3} + 2g_{R\Delta R} + 27g_{RS^2} + 3g_{S^3} \nonumber \\
 \fl & \quad + (-5g_{R^2}^2+540g_{R^3}+20g_{R\Delta R}+27g_{RS^2}+3g_{S^3}) R_k'(z) + 4g_{R^2} R_k''(z) \bigg\}z^3 \nonumber \\
 \fl & \quad - 4\bigg\{ g_{R^2}\big( 56g_{R^2}^2 - 3240g_{R^3} - 300g_{R\Delta R} - 158 g_{RS^2} - 17g_{S^3} \nonumber \\
 \fl & \quad + (2808 g_{R^3} + 104g_{R\Delta R} + 138 g_{RS^2} + 15g_{S^3}) R_k'(z) \big) \nonumber \\
 \fl & \quad + 4(2g_{R^2}^2 - 216g_{R^3} - 8g_{R\Delta R} - 10g_{RS^2} - g_{S^3}) R_k''(z) \bigg\} z^4 \nonumber \\
 \fl & \quad + \bigg\{ -373248g_{R^3}^2 + 1216 g_{R\Delta R}^2 + 912g_{R\Delta R}g_{RS^2} - 992 g_{RS^2}^2 \nonumber \\
 \fl & \quad + 432g_{R^3}(44g_{R\Delta R} - 84g_{RS^2} - 9g_{S^3}) + 96g_{R\Delta R}g_{S^3} - 244g_{RS^2} g_{S^3} - 17g_{S^3}^2 \nonumber \\
 \fl & \quad - 8g_{R^2}^2(3672g_{R^3} + 190g_{R\Delta R} + 178g_{RS^2} + 19g_{S^3}) \nonumber \\
 \fl & \quad - \big( 32(16(27g_{R^3}+g_{R\Delta R})^2+42g_{RS^2}(27g_{R^3}+g_{R\Delta R}) + 31g_{RS^2}^2) \nonumber \\
 \fl & \quad + 4g_{S^3}(972g_{R^3} + 36g_{R\Delta R} + 61g_{RS^2}) + 17g_{S^3}^2 \big) R_k'(z) \nonumber \\
 \fl & \quad - 16g_{R^2}(216 g_{R^3} + 8g_{R\Delta R} + 10g_{RS^2} + g_{S^3}) R_k''(z) \bigg\} z^5 \nonumber \\
 \fl & \quad - 2 \bigg\{ g_{R^2} \big( 466560g_{R^3}^2 + 1648g_{R\Delta R}^2 + 2972g_{R\Delta R}g_{RS^2} + 1224 g_{RS^2}^2 \nonumber \\
 \fl & \quad + 314 g_{R\Delta R} g_{S^3} + 300 g_{RS^2} g_{S^3} + 21g_{S^3}^2 \nonumber \\
 \fl & \quad + 432g_{R^3}(143 g_{R\Delta R} + 104g_{RS^2} + 11g_{S^3}) \big) \nonumber \\
 \fl & \quad + 2 \big( 23328g_{R^3}^2 + 32g_{R\Delta R}^2 + 80g_{R\Delta R} g_{RS^2} + 58 g_{RS^2}^2 + 8 g_{R\Delta R} g_{S^3} \nonumber \\
 \fl & \quad + 14g_{RS^2} g_{S^3} + g_{S^3}^2 + 216g_{R^3}(8g_{R\Delta R}+10g_{RS^2}+g_{S^3}) \big) R_k''(z) \bigg\} z^6 \bigg] \nonumber \\
 \fl & \quad + \frac{1}{72} \int_0^\infty \mu(z) G(z)^5 z^2 (1+2g_{R^2}z+3g_{R\Delta R}z^2 + R_k'(z))^2 \times \nonumber \\
 \fl & \quad \bigg[ 16 - 32g_{R^2}z + (16g_{R^2}^2 - 8(216g_{R^3} + 8g_{R\Delta R} + 10 g_{RS^2} + g_{S^3})) z^2 \nonumber \\
 \fl & \quad + 8g_{R^2} (216g_{R^3} + 8g_{R\Delta R} + 10g_{RS^2} + g_{S^3}) z^3 \nonumber \\
 \fl & \quad + \bigg\{ 17(4g_{RS^2}+g_{S^3})^2 + 8(4g_{RS^2}+g_{S^3})(216g_{R^3} + 8g_{R\Delta R} - 3(2g_{RS^2} + g_{S^3})) \nonumber \\
 \fl & \quad + (216 g_{R^3} + 8g_{R\Delta R} - 3(2g_{RS^2} + g_{S^3}))^2 \bigg\} z^4 \bigg] \, , \\
 \fl\left. \dot \Gamma_k \right|_{C\Delta C} &= \frac{1}{1680} \left( \mu'(0) G(0) + \mu(0) G'(0) \right) \nonumber \\
 \fl & \quad + \frac{1}{360} \int_0^\infty \mu(z) G(z)^2 z (60g_{C\Delta C} + 12g_{RS^2} - 10g_{SSC} + 33g_{S^3}) \nonumber \\
 \fl & \quad + \frac{1}{576} \int_0^\infty \mu(z) G(z)^3 z^4 (4g_{RS^2}+g_{S^3})(108g_{RS^2} + 16g_{SSC} - 21g_{S^3}) \nonumber \\
 \fl & \quad - \frac{1}{96} (4g_{RS^2}+g_{S^3})^2 \int_0^\infty \mu(z) G(z)^4 z^5 \bigg[ 7 + 18g_{R^2}z + 33g_{R\Delta R}z^2 \nonumber \\
 \fl & \quad + 7R_k'(z) + 2z R_k''(z) \bigg] \nonumber \\
 \fl & \quad + \frac{1}{24} (4g_{RS^2} + g_{S^3})^2 \int_0^\infty \mu(z) G(z)^5 z^6 (1+2g_{R^2}z + 3g_{R\Delta R}z^2 + R_k'(z))^2 \, , \\
 \fl\left. \dot \Gamma_k \right|_{R^3} &= - \frac{29}{6480} \mu(0) G(0)^2 - \frac{37}{54432} \left( \mu'(0) G(0) + \mu(0) G'(0) \right) \nonumber \\
 \fl & \quad - \frac{1}{108} \int_0^\infty \mu(z) G(z)^2 z \bigg[ 2(63g_{R^3}+g_{R\Delta R}) + 3 g_{RS^2} \bigg] \nonumber \\
 \fl & \quad + \frac{1}{216} \int_0^\infty \mu(z) G(z)^3 \bigg[ 4 - 8g_{R^2}z + 4 (g_{R^2}^2 + 3(72g_{R^3}+g_{RS^2})) z^2 \nonumber \\
 \fl & \quad - 12 g_{R^2} (72g_{R^3}+g_{RS^2}) z^3 - (108g_{R^3} + 4g_{R\Delta R}+3g_{RS^2}) \times \nonumber \\
 \fl & \quad (540g_{R^3}+4g_{R\Delta R}+9g_{RS^2}) z^4 \bigg] \, , \\
 \fl\left. \dot \Gamma_k \right|_{RS^2} &= - \frac{1}{540} \mu(0) G(0)^2 - \frac{1}{840} \left( \mu'(0) G(0) + \mu(0) G'(0) \right) \nonumber \\
 \fl & \quad - \frac{1}{36} \int_0^\infty \mu(z) G(z)^2 z (2g_{RS^2} + 3 g_{S^3}) \nonumber \\
 \fl & \quad + \frac{1}{288} \int_0^\infty \mu(z) G(z)^3 \bigg[ 64(11g_{RS^2}+2g_{S^3})z^2 - 64g_{R^2}(11g_{RS^2}+2g_{S^3})z^3 \nonumber \\
 \fl & \quad + \bigg\{ -16g_{RS^2}(88(27g_{R^3}+g_{R\Delta R})+113g_{RS^2}) \nonumber \\
 \fl & \quad - 8g_{S^3}(864g_{R^3}+32g_{R\Delta R}+67g_{RS^2}) - 39g_{S^3}^2 \bigg\} z^4 \bigg] \nonumber \\
 \fl & \quad + \frac{1}{48} \int_0^\infty \mu(z) G(z)^4 (4g_{RS^2}+g_{S^3})^2 \bigg[ (216g_{R^3} + 41g_{R\Delta R}+6g_{RS^2}) z^7 \nonumber \\
 \fl & \quad + z^5(3+7 R_k'(z)) + 2z^6(11g_{R^2}+R_k''(z)) \bigg] \nonumber \\
 \fl & \quad - \frac{1}{12} \int_0^\infty \mu(z) G(z)^5 (4g_{RS^2}+g_{S^3})^2 \bigg[ 12g_{R^2} g_{R\Delta R} z^9 + 9g_{R\Delta R}^2 z^{10} \nonumber \\
 \fl & \quad + 4g_{R^2} z^7 (1+R_k'(z)) + z^6(1+R_k'(z))^2+2z^8(2g_{R^2}^2 + 3g_{R\Delta R}(1+R_k'(z))) \bigg] \, , \\
 \fl\left. \dot \Gamma_k \right|_{S^3} &= - \frac{1}{1890} \left( \mu'(0) G(0) + \mu(0) G'(0) \right) + \frac{1}{2} g_{S^3} \int_0^\infty \mu(z) G(z)^2 z \nonumber \\
 \fl & \quad - \frac{1}{48} \int_0^\infty \mu(z) G(z)^3 z^4 (4g_{RS^2}+g_{S^3})(172g_{RS^2}+51g_{S^3}) \nonumber \\
 \fl & \quad + \frac{1}{24} \int_0^\infty \mu(z) G(z)^4 (4g_{RS^2}+g_{S^3})^2 \bigg[ (99g_{R\Delta R}+4g_{RS^2}+g_{S^3})z^7 \nonumber \\
 \fl & \quad + 21z^5(1+R_k'(z)) + 6z^6(9g_{R^2}+R_k''(z)) \bigg] \nonumber \\
 \fl & \quad - \frac{1}{2} \int_0^\infty \mu(z) G(z)^5 (4g_{RS^2}+g_{S^3})^2 \bigg[ 12g_{R^2} g_{R\Delta R} z^9+9g_{R\Delta R}^2 z^{10} \nonumber \\
 \fl & \quad +4g_{R^2}z^7(1+R_k'(z))+z^6(1+R_k'(z))^2 \nonumber \\
 \fl & \quad + 2z^8(2g_{R^2}^2+3g_{R\Delta R}(1+R_k'(z))) \bigg] \, , \\
 \fl\left. \dot \Gamma_k \right|_{SSC} &= - \frac{1}{420} \left( \mu'(0) G(0) + \mu(0) G'(0) \right) \nonumber \\
 \fl & \quad + \frac{1}{180} (-12g_{RS^2}+70g_{SSC}+27g_{S^3}) \int_0^\infty \mu(z) G(z)^2 z \nonumber \\
 \fl & \quad + \frac{1}{288} \int_0^\infty \mu(z) G(z)^3 z^4 (4g_{RS^2}+g_{S^3}) \bigg[ 516g_{RS^2}-16g_{SSC}+81g_{S^3} \bigg] \nonumber \\
 \fl & \quad -\frac{1}{16} \int_0^\infty \mu(z) G(z)^4 (4g_{RS^2}+g_{S^3})^2 \bigg[ 33g_{R\Delta R}z^7 \nonumber \\
 \fl & \quad + 7z^5(1+R_k'(z))+2z^6(9g_{R^2}+R_k''(z)) \bigg] \nonumber \\
 \fl & \quad + \int_0^\infty \mu(z) G(z)^5 (4g_{RS^2}+g_{S^3})^2 \bigg[ 12g_{R^2}g_{R\Delta R} z^9 + 9g_{R\Delta R}^2 z^{10} \nonumber \\
 \fl & \quad + 4g_{R^2}z^7(1+R_k'(z)) + z^6(1+R_k'(z))^2 \nonumber \\
 \fl & \quad + 2z^8(2g_{R^2}^2+3g_{R\Delta R}(1+R_k'(z))) \bigg] \, , \\
 \fl\left. \dot \Gamma_k \right|_{RC^2} &=  - \frac{1}{540} \mu(0) G(0)^2 - \frac{1}{1080} \left( \mu'(0) G(0) + \mu(0) G'(0) \right) \nonumber \\
 \fl & \quad + \frac{1}{720} \int_0^\infty \mu(z) G(z)^2 z \bigg[ 100 g_{C\Delta C} - 120 g_{RC^2} + 12 g_{RS^2} + 33 g_{S^3} \bigg] \nonumber \\
 \fl & \quad + \frac{1}{1152} \int_0^\infty \mu(z) G(z)^3 \bigg[ -64(10g_{C\Delta C}-36g_{RC^2}+g_{SSC})z^2 \nonumber \\
 \fl & \quad + 64g_{R^2}(10g_{C\Delta C}-36g_{RC^2}+g_{SSC})z^3 + \bigg\{ 320g_{C\Delta C}(108g_{R^3}+4g_{R\Delta R}+3g_{RS^2}) \nonumber \\
 \fl & \quad -144(32g_{RC^2}(27g_{R^3}+g_{R\Delta R})+24g_{RC^2}g_{RS^2}-3g_{RS^2}^2) \nonumber \\
 \fl & \quad + 32g_{SSC}(108g_{R^3}+4g_{R\Delta R}+5g_{RS^2}) + 8g_{S^3}(3g_{RS^2}+2g_{SSC}) - 21g_{S^3}^2 \bigg\} z^4 \bigg] \nonumber \\
 \fl & \quad - \frac{1}{192} \int_0^\infty \mu(z) G(z)^4 (4g_{RS^2}+g_{S^3})^2 \bigg[ 33g_{R\Delta R}z^7 + 7z^5(1+R_k'(z)) \nonumber \\
 \fl & \quad + 2z^6(9g_{R^2}+R_k''(z)) \bigg] \nonumber \\
 \fl & \quad + \frac{1}{48} \int_0^\infty \mu(z) G(z)^5 (4g_{RS^2}+g_{S^3})^2 \bigg[ 12 g_{R^2} g_{R\Delta R} z^9 + 9g_{R\Delta R}^2 z^{10} \nonumber \\
 \fl & \quad + 4g_{R^2} z^7(1+R_k'(z)) + z^6(1+R_k'(z))^2 + 2z^8(2g_{R^2}^2+3g_{R\Delta R}(1+R_k'(z))) \bigg] \, , \\
 \fl\left. \dot \Gamma_k \right|_{C^3} &= - \frac{1}{540} \left( \mu'(0) G(0) + \mu(0) G'(0) \right) \nonumber \\
 \fl & \quad + \frac{1}{120} \int_0^\infty \mu(z) G(z)^2 z \bigg[ 40 g_{C^3} - 12 g_{RS^2} + 10 g_{SSC} - 33 g_{S^3} \bigg] \nonumber \\
 \fl & \quad + \frac{1}{192} \int_0^\infty \mu(z) G(z)^3 z^4 \bigg[ (4g_{RS^2} + g_{S^3})(-108g_{RS^2}-16g_{SSC}+21g_{S^3}) \bigg] \nonumber \\
 \fl & \quad + \frac{1}{32} \int_0^\infty \mu(z) G(z)^4 (4g_{RS^2}+g_{S^3})^2 \bigg[ 33g_{R\Delta R} z^7 + 7 z^5 (1+R_k'(z)) \nonumber \\
 \fl & \quad + 2 z^6 (9g_{R^2}+R_k''(z)) \bigg] \nonumber \\
 \fl & \quad - \frac{1}{8} \int_0^\infty \mu(z) G(z)^5 (4g_{RS^2}+g_{S^3})^2 \bigg[ 12 g_{R^2} g_{R\Delta R} z^9 + 9g_{R\Delta R}^2z^{10} + z^6 (1+R_k'(z))^2 \nonumber \\
 \fl & \quad + 4g_{R^2} z^7 (1+R_k'(z)) + 2z^8 (2g_{R^2}^2+3g_{R\Delta R}(1+R_k'(z))) \bigg] \, .
\end{eqnarray}
Notice that here we already formulated everything in the traceless tensor basis. In particular, the term corresponding to the $R^2$ operator is already the flow of the respective coupling. For completeness, we also list the flow of the two remaining operators at this order:
\begin{eqnarray}
 \fl\left. \dot \Gamma_k \right|_{C^2} &= \frac{1}{120} \mu(0) G(0) + \frac{1}{12} \int_0^\infty \mu(z) G(z)^2 z^2 \bigg[ 10g_{C\Delta C} - 36g_{RC^2} - 22g_{RS^2} + g_{SSC} - 4g_{S^3} \bigg] \nonumber \\
 \fl &\quad + \frac{1}{24} (4g_{RS^2} + g_{S^3})^2 \int_0^\infty \mu(z) G(z)^3 z^5 \, , \\
 \fl\left. \dot \Gamma_k \right|_{\mathfrak E} &= -\frac{1}{360} \mu(0) G(0) + \frac{1}{6} (11g_{RS^2} + 2g_{S^3}) \int_0^\infty \mu(z) G(z)^2 z^2 \nonumber \\
 \fl &\quad - \frac{1}{24} (4g_{RS^2} + g_{S^3})^2 \int_0^\infty \mu(z) G(z)^3 z^5 \, .
\end{eqnarray}
A generic feature of such traces based on the early-time heat kernel expansion is that as soon as the mass dimension of a tensor structure reaches the spacetime dimension, non-integral terms arise. These come from terms in the trace with non-negative powers of the heat parameter. Characteristically, these contributions are (derivatives of) an integrand evaluated at zero.

\section{Basis for operators of \texorpdfstring{$\mathcal O(\partial^8)$}{order 8}}\label{app:basis}

For future reference, we provide a basis for the operators with eight derivatives. This choice of basis is inspired by \cite{Fulling:1992vm, 2007CoPhC.177..640M, 2008CoPhC.179..586M} and adapted to traceless tensors. The full basis reads
\begin{eqnarray}
\fl \mathcal L_8 &= -\frac{1}{6} g_{R\Delta^2R} R \Delta^2 R + \frac{1}{2} g_{C\Delta^2C} C^{\mu\nu\rho\sigma} \Delta^2 C_{\mu\nu\rho\sigma} + g_{RDRDR} R (D^\mu R) (D_\mu R) \nonumber \\
 \fl &\quad + g_{SDRDR} S^{\mu\nu} (D_\mu R) (D_\nu R) + g_{RDSDS_1} R (D^\rho S^{\mu\nu}) (D_\rho S_{\mu\nu}) \nonumber \\
 \fl &\quad + g_{RDSDS_2} R (D^\rho S^{\mu\nu}) (D_\mu S_{\nu\rho}) + g_{SDSDR} S^{\nu\rho} (D_\mu R) (D_\mu S_{\nu\rho}) \nonumber \\
 \fl &\quad + g_{SDSDS_1} S^{\mu\nu} (D_\sigma S_{\mu\rho}) (D^\sigma S_{\nu}^{\phantom{\nu}\rho}) + g_{SDSDS_2} S^{\mu\nu} (D^\sigma S_{\mu\rho}) (D^\rho S_{\nu\sigma}) \nonumber \\
 \fl & \quad + g_{SDSDS_3} S^{\mu\nu} ( D_\mu S_{\rho\sigma} ) ( D_\nu S^{\rho\sigma} ) + g_{SDSDS_4} S^{\mu\nu} (D_\mu S_{\rho\sigma} ) (D^\sigma S_\nu^{\phantom{\nu}\rho}) \nonumber \\
 \fl & \quad + g_{CDRDS} C_{\mu\nu\rho\sigma} (D^\mu R) (D^\sigma S^{\nu\rho}) + g_{CDSDS_1} C^{\nu\rho\sigma\alpha} (D_\rho S_{\mu\nu}) (D_\alpha S^\mu_{\phantom{\mu}\sigma}) \nonumber \\
 \fl & \quad + g_{CDSDS_2} C_{\nu\sigma\rho\alpha} (D^\mu S^{\nu\rho}) (D_\mu S^{\sigma\alpha}) + g_{CDSDS_3} C_{\mu\sigma\rho\alpha} (D^\rho S^{\mu\nu}) (D_\mu S^{\sigma\alpha}) \nonumber \\
 \fl & \quad + g_{SDCDS} S^{\mu\nu} (D^\alpha S^{\rho\sigma}) (D_\alpha C_{\mu\rho\nu\sigma}) + g_{CDCDR} C^{\nu\rho\sigma\alpha} (D^\mu R) (D_\mu C_{\nu\rho\sigma\alpha}) \nonumber \\
 \fl & \quad + g_{RDCDC} R (D^\mu C^{\nu\rho\sigma\alpha}) (D_\mu C_{\nu\rho\sigma\alpha}) + g_{SDCDC} S^{\alpha\beta} (D_\rho C_{\beta\delta\gamma\sigma}) (D^\sigma C_\alpha^{\phantom{\alpha}\gamma\delta\rho}) \nonumber \\
 \fl & \quad + g_{CDCDS} C_{\mu\alpha\beta\gamma} (D^\rho S^{\mu\nu}) (D_\nu C_\rho^{\phantom{\rho}\alpha\beta\gamma}) + g_{CDCDC} C^{\mu\nu\rho\sigma} (D_\alpha C_{\rho\sigma\beta\gamma}) (D^\alpha C_{\mu\nu}^{\phantom{\mu\nu}\beta\gamma}) \nonumber \\
 \fl & \quad + g_{R^4} R^4 + g_{R^2S^2} R^2 S^{\mu\nu} S_{\mu\nu} + g_{RS^3} R S^\mu_{\phantom{\mu}\nu} S^\nu_{\phantom{\nu}\rho} S^\rho_{\phantom{\rho}\mu} + g_{S^2S^2} S^{\mu\nu} S_{\mu\nu} S^{\rho\sigma} S_{\rho\sigma} \nonumber \\
 \fl & \quad + g_{S^4} S^\mu_{\phantom{\mu}\nu} S^\nu_{\phantom{\nu}\rho} S^\rho_{\phantom{\rho}\sigma} S^\sigma_{\phantom{\sigma}\mu} + g_{RSSC} R S^{\mu\nu} S^{\rho\sigma} C_{\mu\rho\nu\sigma} + g_{R^2C^2} R^2 C^{\mu\nu\rho\sigma} C_{\mu\nu\rho\sigma} \nonumber \\
 \fl & \quad + g_{S^2C^2} S^{\alpha\beta} S_{\alpha\beta} C^{\mu\nu\rho\sigma} C_{\mu\nu\rho\sigma} + g_{SSCC_1} S^{\mu\nu} S^{\rho\sigma} C^{\alpha\beta}_{\phantom{\alpha\beta}\mu\rho} C_{\alpha\beta\nu\sigma} \nonumber \\
 \fl & \quad + g_{SSCC_2} S^{\mu\nu} S^{\rho\sigma} C^{\alpha\phantom{\mu}\beta}_{\phantom{\alpha}\mu\phantom{\beta}\nu} C_{\alpha\rho\beta\sigma} + g_{RC^3} R C^{\mu\nu}_{\phantom{\mu\nu}\rho\sigma} C^{\rho\sigma}_{\phantom{\rho\sigma}\tau\omega} C^{\tau\omega}_{\phantom{\tau\omega}\mu\nu} \nonumber \\
 \fl & \quad + g_{C^2C^2} C^{\mu\nu\rho\sigma} C_{\mu\nu\rho\sigma} C^{\alpha\beta\gamma\delta} C_{\alpha\beta\gamma\delta} + g_{C^4} C^{\mu\nu\rho\sigma} C_{\mu\nu}^{\phantom{\mu\nu}\tau\omega} C_{\tau\omega}^{\phantom{\tau\omega}\kappa\lambda} C_{\rho\sigma\kappa\lambda} \, .
\end{eqnarray}
From the sheer number of couplings, it is clear that an analysis at this order will be extremely challenging.

~

\section*{References}

\bibliographystyle{iopart-num}
\bibliography{general_bib.bib}

\end{document}